\documentclass[letterpaper,journal]{IEEEtran}
\usepackage{amsmath,amsfonts}
\usepackage{algorithmic}
\usepackage{algorithm}
\usepackage{array}
\usepackage{textcomp}
\usepackage{stfloats}
\usepackage{url}
\usepackage{verbatim}
\usepackage{graphicx}
\usepackage{cite}
\usepackage{booktabs}
\usepackage{makecell}
\usepackage{multirow}
\usepackage{subcaption}
\begin{document}

\title{Graph-Based ECO and Patch Generation for High-Level Synthesis}

\author{Alireza~Azadi,
        Paul~Rigge,
        Ethan~Mahintorabi,
        and Kenneth~B.~Kent%
\thanks{Alireza Azadi and Kenneth B. Kent are with the Faculty of Computer Science, 
University of New Brunswick, Fredericton, NB, Canada 
(e-mail: al.azadi94@unb.ca; ken@unb.ca).}%
\thanks{Paul Rigge and Ethan Mahintorabi are with Google, Mountain View, CA, USA 
(e-mail: rigge@google.com; ethanmoon@google.com).}%
\thanks{The source code and test cases are available at 
\protect\url{https://github.com/alirezazd/xls-eco/tree/eco-paper-2025}.}}

% The paper headers
\markboth{IEEE Transactions on Computer-Aided Design of Integrated Circuits and Systems}%
{Azadi \MakeLowercase{\textit{et al.}}: Graph-Based ECO and Patch Generation for High-Level Synthesis}

% \IEEEpubid{Will be assigned by IEEE upon acceptance}

\maketitle

\begin{abstract}
	High-level synthesis (HLS) tools offer limited support for Engineering
	Change Orders (ECOs), making late-stage design modifications challenging
	and costly. This paper introduces a graph-based ECO methodology tailored
	for Google XLS. A Graph Edit Distance (GED) algorithm is used to detect
	structural differences between original and revised intermediate
	representations (IRs), which are then transformed into
	patch operations. A patch application mechanism is developed to enforce XLS IR
	constraints while preserving semantic correctness, together with a schedule
	constraining scheme that maintains the original pipeline registers.
	Experiments across several XLS designs demonstrate high structural reuse ratios,
	effective schedule preservation, and full functional correctness,
	highlighting the practicality of the approach for production HLS flows.
\end{abstract}


\section{Introduction}
\IEEEPARstart{E}{ngineering} Change Orders (ECOs) play a critical role in enabling late-stage
modifications to hardware designs, particularly in ASIC development, where the
cost and time associated with late-stage changes are significant. While
traditional ECO methods focus primarily on Register-Transfer Level (RTL)
designs, they often fall short in the context of High-Level Synthesis (HLS), where
small high-level changes can drastically alter output RTL due to heavy
optimizations, compiler transformations, and scheduling variations inherent in
HLS flows.

Google XLS, the focus of this paper, is an open-source HLS toolchain featuring
frontends that transform high-level code into optimized and scheduled IR.
However, minor high-level modifications can produce unpredictable effects on
the post-optimized IR due to cascading optimization and scheduling
transformations. This unpredictability makes it difficult to trace change
impacts, and since RTL is directly derived from the IR, the absence of
HLS-level ECO capabilities limits effective change management throughout the
design flow, ultimately constraining RTL-level ECO and compromising design
flexibility at the chip level.

\begin{figure}[!t]
	\centering
	\subfloat[Original IR]{\includegraphics[height=3.3cm]{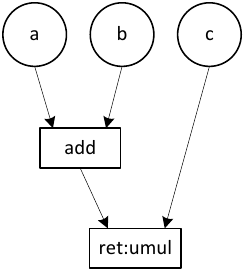}%
		\label{fig:eco_org}}
	\hfil
	\subfloat[Patched IR]{\includegraphics[height=3.3cm]{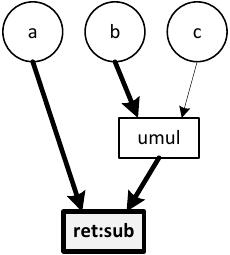}%
		\label{fig:eco_rev}}

	\vspace{0.4cm}

	\subfloat[Patch Operations]{\includegraphics[height=2.2cm]{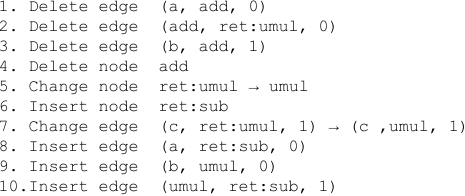}%
		\label{fig:eco_patch}}
	\caption{Illustration of IR patching: (a) original IR implementing \texttt{ret = (a + b) * c}; (b) patched IR implementing \texttt{ret = a - (b * c)}; (c) patch operations generated by the diff tool. Newly inserted nodes and edges are highlighted with boldness and gray fill. Reused elements are preserved without change.}
	\label{fig:eco_overview}
\end{figure}

To demonstrate the core concept of our approach, we present a minimal example
in Figure~\ref{fig:eco_overview}. Starting from a simple arithmetic IR
implementing \texttt{ret = (a + b) * c}, we modify it to \texttt{ret = a - (b *
	c)}. This seemingly small change requires structural edits at the IR level,
including deletion and insertion of nodes and edges. Our ECO flow identifies
these differences, generates a compact patch, and applies it while reusing
unchanged parts of the original IR. The patch listing summarizes the precise
edit operations computed by the tool. \IEEEpubidadjcol

This paper proposes a novel graph-based ECO flow specifically targeting XLS IR.
The methodology utilizes Graph Edit Distance (GED) techniques to calculate
minimal transformation paths between the original and modified IRs. These paths
are translated into patch operations, ensuring that changes are applied while
respecting the integrity of the original design. Key features of the proposed
approach include a GED-based diffing tool that incorporates XLS-aware cost
metrics, a patch application mechanism that maintains structural correctness,
and a schedule-constraining scheme to preserve timing alignment. The approach
is evaluated on several real-world XLS designs, demonstrating high structural
reuse (up to 95\%), effective schedule preservation (up to 92\%), and
correctness through functional equivalence verification.

The contributions of this work are as follows:
\begin{itemize}
	\item A complete end-to-end ECO methodology tailored for XLS IR, offering a
	      systematic solution to address the challenges of late-stage design
	      modifications.
	\item A GED-based graph differencing tool that incorporates structural semantics,
	      ensuring precise and effective IR transformations.
	\item A robust patch generation and application mechanism that minimizes disruptions
	      to the original design, ensuring structural and semantic correctness.
	\item A schedule-preserving approach that retains cycle-level timing and ensures RTL
	      timing alignment in the final design.
	\item Extensive evaluation on diverse XLS designs, demonstrating high reuse ratios,
	      minimal structural changes, and efficient runtime performance.
\end{itemize}
Our complete ECO methodology is fully implemented in an open-source flow based
on Google’s XLS framework~\cite{xls_eco_repo}. The rest of the paper is
organized as follows: Section \ref{sec:background} provides background on
high-level synthesis, the XLS framework, and GED algorithms. Section
\ref{sec:related_work} reviews related work in ECO methodologies at both the
high-level and RTL levels. Section \ref{sec:methodology} outlines the proposed
ECO flow and its components. Section \ref{sec:results} presents the
experimental setup, results, and performance analysis of the proposed
methodology. Finally, Section \ref{sec:conclusion} concludes the paper and
suggests directions for the future work.
\section{Background}
\label{sec:background}
This section provides essential background on the key technologies and concepts
underlying the ECO methodology, including an overview of High-Level Synthesis
and its role in modern hardware design, a description of Google XLS and its IR,
and, finally, Graph Edit Distance as the theoretical foundation of the approach.
\subsection{High-Level Synthesis (HLS)}
As indicated by Moore’s Law~\cite{Moore1998} increasing transistor density
inherently drives greater hardware design complexity. Initially, hardware
designers relied on RTL languages like Verilog and VHDL, which provided a
structured way to describe digital circuits but still required meticulous
low-level design efforts. However, as designs grew more complex, the need for
higher levels of abstraction became apparent. High-Level Synthesis (HLS) has
been a major focus of CAD researchers since the late 1970s~\cite{HLS_PPF},
aiming to bridge the gap between algorithmic descriptions and hardware
implementation. By the 1990s, the first generation of commercial HLS tools
became available~\cite{intro_HLS}, marking a shift toward automation in
hardware design. HLS not only abstracts hardware description but also
facilitates system-level exploration, enabling architectural decisions related
to hardware-software partitioning, memory organization, and power management.
Like the relationship between assembly language and high-level programming, RTL
provides precise control over hardware optimizations, while HLS enables a more
abstract design flow that accelerates development. Though not as optimized as
handcrafted RTL, HLS automates tasks such as concurrency management, pipeline
insertion and control logic synthesis~\cite{PPFPGAs}.
\subsection{Google XLS}
Google XLS (Accelerated HW Synthesis)~\cite{google_xls} is an open-source
hardware design toolchain used both internally at Google and by the wider
research community~\cite{TFHE, antmicro_xls, xls_delay}. XLS provides a
flexible intermediate representation (IR) tailored for dataflow computations.
The general flow of XLS involves describing hardware at a high level in DSLX or
C++, which is then converted to IR. The IR undergoes multiple optimization
passes such as arithmetic simplification and conditional specializations before
producing a post-optimized IR. The IR is then scheduled into a pipelined
implementation, from which register-transfer level (RTL) code is generated
\subsection{Graph Edit Distance (GED) Algorithms}
Graph Edit Distance (GED) is a fundamental metric for quantifying the
dissimilarity between two graphs. It is defined as the minimum-cost sequence of
operations, namely, node and edge insertions, deletions, or substitution
required to transform one graph into another. Formally, given two graphs $G_1$
and $G_2$, the GED is defined as:
\begin{equation}
	\label{eq:ged_def}
	\text{GED}(G_1, G_2) = \min_{P \in \mathcal{P}(G_1, G_2)} \sum_{o \in P} \text{cost}(o)
\end{equation}
where $\mathcal{P}(G_1, G_2)$ is the set of all valid edit paths transforming
$G_1$ into $G_2$, and $\text{cost}(o)$ denotes the cost of an individual edit
operation $o$. While GED yields a scalar distance, Graph Edit Paths (GEPs)
extend this notion by explicitly returning the sequence of edit operations,
enabling interpretability in transformation scenarios. GED is widely adopted in
domains such as pattern recognition, bioinformatics, and circuit optimization,
where both similarity assessment and precise structural modifications are
essential. Abu-Aisheh et al.~\cite{AbuAisheh2015} propose DF-GED, an exact GED
algorithm based on depth-first branch-and-bound search. The algorithm leverages
precomputed vertex and edge cost matrices as well as a Munkres-based sorting
~\cite{kuhn1955hungarian} of source nodes to prioritize promising mappings
early. While the worst-case time complexity remains exponential, DF-GED
demonstrates significant empirical gains in both runtime and memory efficiency.
The DF-GED algorithm is implemented as part of the NetworkX
library~\cite{networkx_optimize_edit_paths} and we found it to be a good
candidate to demonstrate our methodology in this paper due to the flexibility
of NetworkX which allows us to represent our IR as a graph and the exact output
that is mandatory for the functional correctness of ECO.

\section{Related Work}
\label{sec:related_work}

Although ECO methodologies have been extensively studied at the RTL and gate
levels, only a limited academic efforts study this important challenge for
High-Level Synthesis (HLS) ~\cite{lavagno2010incremental, datapath_eco_2019}.
ECO at the HLS level poses unique difficulties due to the higher level of
abstraction, where even minor source changes can trigger widespread and
unpredictable transformations in the resulting
netlist~\cite{lavagno2010incremental}. This complicates change localization and
makes the design of scalable patching mechanisms more difficult.

In this section, we review prior ECO methodologies from HLS to gate level,
focusing on patch generation, scheduling preservation, and the availability of
supporting tools. A comparative summary is provided in
Table~\ref{tab:eco-comparison}, which includes our proposed graph-based ECO
flow. This comparison underscores the key characteristics of our approach,
including fine-grained patching at the graph level, robust schedule
preservation and full integration within an open-source flow, while also
identifying areas where prior work leaves room for improvement.
\begin{table*}[!t]
	\centering
	\caption{Comparison of ECO methodologies against our graph-based HLS flow}
	\label{tab:eco-comparison}
	\footnotesize
	\renewcommand{\arraystretch}{1.3}
	\begin{tabular}{|l l l l l l|}
		\hline
		\textbf{Work}                                   & \textbf{Target Level} & \textbf{Technique}                                                               & \textbf{Patch Granularity} & \textbf{Schedule Preservation} & \textbf{Open Source} \\
		\hline
		\textbf{This Work}                              & HLS                   & GED                                                                              & Individual Nodes \& Edges  & Yes                            & Yes                  \\
		Alizadeh et al.~\cite{datapath_eco_2019}        & HLS                   & SMT                                                                              & Connections \& Muxes       & Yes                            & No                   \\
		Lavagno et al.~\cite{lavagno2010incremental}    & HLS                   & String Diff                                                                      & Operator-Level             & Partial                        & No                   \\
		Wang et al.~\cite{wang2019programmable}         & HLS                   & \begin{tabular}[t]{@{}l@{}}Programmable Datapath\\+ SMT/QBF\end{tabular}         & Operator-Level             & No                             & No                   \\
		Alizadeh et al.~\cite{Automatic_RTL_Correction} & RTL                   & Subgraph Match                                                                   & Region-level               & No                             & No                   \\
		Zhang et al.~\cite{zhang2018costaware}          & Gate                  & QBF-guided Rectification                                                         & Logic Cone                 & N/A                            & No                   \\
		Dao et al.~\cite{dao2018efficient}              & Gate                  & SAT-based Patch Synthesis                                                        & Logic Cone                 & N/A                            & No                   \\
		Cheng et al.~\cite{cheng2016resourceaware}      & Gate                  & \begin{tabular}[t]{@{}l@{}}Craig Interpolation\\+ Greedy Heuristics\end{tabular} & Logic Cone                 & N/A                            & No                   \\
		Kravets et al.~\cite{kravets2019symbolic}       & Gate                  & SAT + Symbolic Sampling                                                          & Logic Cone                 & N/A                            & No                   \\
		Krishnaswamy et al.~\cite{deltasyn2009}         & Gate                  & Structural Hashing + SAT                                                         & Logic Cone                 & N/A                            & No                   \\
		\hline
	\end{tabular}
\end{table*}
\subsection{High-Level ECO Methods}
Post-synthesis ECO methods operate on gate-level netlists. At the gate level,
operation-level scheduling and pipeline timing constraints are no longer
explicitly available and are resolved earlier, during high-level or RTL
synthesis. As a result, post-HLS ECO techniques focus on functional correctness
and structural patching, rather than preserving operation schedules. Alizadeh
et al.~\cite{datapath_eco_2019} propose a datapath-aware ECO synthesis method
targeting the high-level synthesis stage. Their approach aims to preserve the
original datapath structure by reusing existing functional units and registers.
To achieve this goal, they selectively modify the scheduling and binding phases
and utilize an SMT-based formulation to minimize changes to connections and
multiplexers. It is worth noting that their methodology is implemented within a
custom toolchain, and its compatibility with commercial or open-source HLS
flows has not been demonstrated. Lavagno et al.~\cite{lavagno2010incremental}
propose an incremental high-level synthesis approach to enable ECO support by
reusing scheduling and binding decisions across synthesis runs. Their
methodology operates on SystemC inputs and internal Control-Data Flow Graph
(CDFG) representations. Similar to our approach, they utilize diffing
techniques for change detection; however, their method relies on string-based
differencing of linearized graphs, whereas our approach employs graph-based
differencing. Their patching is at the operation level (nodes), and their
scheduling preservation is partial: prior decisions are reused if valid, but
otherwise discarded. Their approach, implemented within a proprietary synthesis
framework, lacks robustness when small source changes introduce new
constraints. If prior scheduling decisions become invalid, the tool discards
them entirely instead of selectively preserving what remains feasible. Wang et
al.~\cite{wang2019programmable} propose a methodology for ECO centered around
use of programmable datapaths, by formulating equivalence and leveraging SMT
solving using Z3~\cite{Z3} to derive valid datapath configurations. Patch
generation is performed by rerouting through spare operators using multiplexer
control signals, making their patch scope operator-level. This method
demonstrates potential for small-scale designs, however, their approach
struggles with larger datapaths due to solver complexity. In terms of
reproducibility, their work is based on a custom flow that is not integrated
into any open-source or commercial HLS framework. Unlike our ECO methodology,
which integrates a schedule-constraining phase to preserve RTL timing
alignment, Wang et al.’s method does not address scheduling or pipeline
preservation. Their focus remains at the datapath control level, assuming
structural reuse will inherently maintain timing; a limitation in sequential or
deeply pipelined designs.
\subsection{Post-HLS ECO}
In a subsequent study~\cite{Automatic_RTL_Correction}, Alizadeh et al. shift
their focus to the RTL level, proposing a method that synthesizes localized
patches by extracting subgraphs from the RTL and a C/C++ reference. Their patch
granularity is region-level, targeting entire datapath structures. Their
methodology emphasizes datapath preservation and as a result, it may overlook
updates to combinational logic, such as those introduced by pipelining
optimizations. Zhang et al.’s work on cost-aware RTL ECO patch
generation~\cite{zhang2018costaware} uses Quantified Boolean Formula (QBF)
reasoning to find minimal sets of input signals (support) needed to generate
gate-level patches that are functionally correct and efficient in area. The
generated patches target logic cones, meaning that each patch replaces or
rewires the entire logic required to compute a target signal, rather than
modifying fine-grained gates. While the proposed technique effectively targets
structurally conservative ECOs, it operates within a closed academic toolchain
with limited reproducibility. Dao et al.~\cite{dao2018efficient} propose a
SAT-based ECO methodology targeting post-synthesis gate-level netlists. Their
flow synthesizes logic patch functions using cube enumeration and QBF
enhancements. Like Zhang et al.'s approach, their patches operate at the logic
cone level. While functionally robust, their approach assumes that structural
reuse is sufficient for timing preservation and does not address scheduling or
pipeline alignment. Dao et al.’s work is reproducible to some extent for
experts familiar with ABC and SAT-based flows, but the absence of released
scripts and tool modifications imposes significant engineering effort to
replicate. Cheng et al.~\cite{cheng2016resourceaware} present a resource-aware
ECO methodology that improves patch feasibility by selecting support signals
based on physical and structural accessibility within the gate-level netlist.
Their approach integrates Craig interpolation with greedy support selection
heuristics. Like other gate-level approaches, their patches operate at the
logic cone level. Kravets et al.~\cite{kravets2019symbolic} present a robust
ECO rectification methodology for gate-level designs, where structural
dissimilarity between the optimized implementation and the revised
specification complicates patch generation. Their approach uses symbolic
sampling and SAT-based reasoning to identify rectification points and
synthesize minimal-impact rewiring patches. While highly effective for
post-synthesis, their methodology is not integrated with open-source
toolchains, which limits reproducibility. Krishnaswamy et al.'s work on
DeltaSyn~\cite{deltasyn2009} targets gate-level ECO by identifying and
optimizing logic differences between the original synthesized netlist and a new
netlist derived from revised RTL. DeltaSyn is presented as a research prototype
and combines Boolean subcircuit matching, structural hashing, and SAT-based
validation to detect changed logic cones and generate minimal-area patches.
While DeltaSyn is not directly reproducible due to the absence of source code
and tooling details, the conceptual flow reasonably well-described and can be
replicated by experts in the field.

\subsection{Commercial Tools}
Several commercial HLS tools have introduced incremental and ECO-aware
capabilities, although their scope and transparency vary significantly. Cadence
Stratus HLS offers the most direct support for HLS-level ECO flows, featuring a
dedicated ECO mode that prioritizes RTL similarity over re-optimization when
minor high-level changes occur. This mode is tightly integrated with Cadence’s
Conformal ECO and logic synthesis tools with support for bidirectional mapping
between RTL and the source code~\cite{cadence_eco_blog, cadence_stratus_pr}.
Siemens Catapult HLS supports incremental compilation through caching
scheduling and binding results across runs. When applied correctly, this
preserves the micro-architecture of unchanged modules, leading to stable RTL
and behavior under minor C++/SystemC edits~\cite{edn_catapult8}. Xilinx Vitis
HLS and Intel HLS Compiler enable ECO-like flows primarily through downstream
reuse via hierarchical partitioning and incremental
compilation.~\cite{intel_rr,xilinx_incremental}.

\section{Methodology}
\label{sec:methodology}

\begin{figure}[!t]
	\centering
	\includegraphics[width=\linewidth]{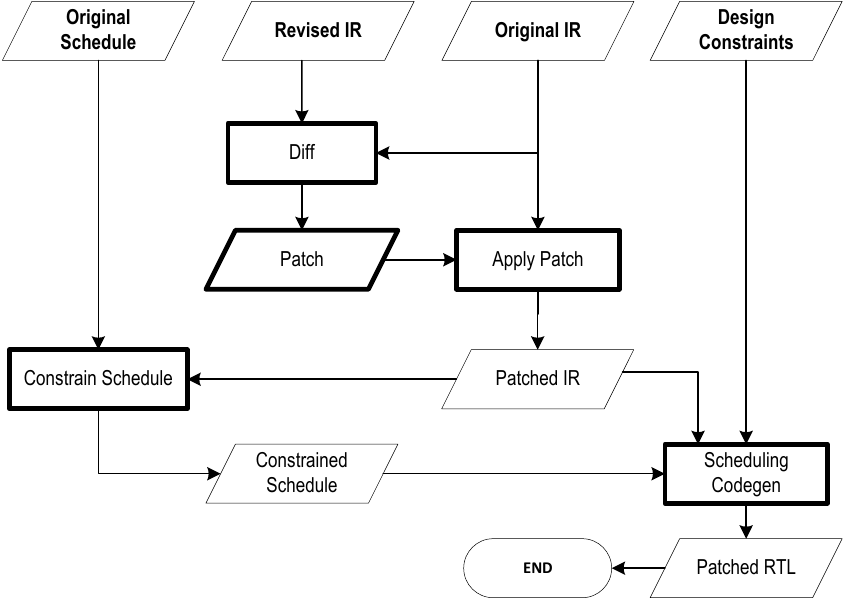}
	\caption{Overview of the XLS ECO Flow. Components with bold borders represent the proposed modules that are integrated into the original XLS flow.}

	\label{fig:eco_flow}
\end{figure}

The flowchart in Figure~\ref{fig:eco_flow} illustrates the systematic process
of applying and validating an ECO to an IR in XLS. The process begins with the
comparison of the ``Original IR'' and ``Revised IR'' through the ``Diff'' tool,
which subsequently generates a ``Patch.'' This patch is then applied to the
original IR, producing a ``Patched IR.'' The flow then proceeds to constrain
the schedule, ensuring that the original timing constraints are preserved. The
final phase involves integrating the constrained schedule, design constraints,
and the patched IR into the ``Codegen'' process, which produces the ``Patched
RTL.''

\subsection{IR to NetworkX Parsing}
The initial step in the ECO flow is translating the XLS IR into a NetworkX
MultiDiGraph representation. We developed an IR parser that converts XLS IR
structures into graph representations suitable for graph-based analysis. The
MultiDiGraph is well-suited for representing XLS IRs because:
\begin{itemize}
	\item It allows multiple parallel edges between the same nodes, which is essential
	      for representing instances where multiple signals may connect the same pair of
	      nodes. For example, a select node with multiple inputs producing the same
	      constant value. This requires multiple edges from the constant literal node to
	      the select node.
	\item It inherently models the one-way flow of data between nodes, reflecting the
	      static single-assignment (SSA) property and the lack of sequential control flow
	      in XLS IRs.
\end{itemize}
Additionally, our IR parser extracts the attributes of the nodes
and edges in the graph. These attributes are later utilized by the diff tool.
\subsection{IR Diff Tool}
The IR diffing tool is a core component of the ECO flow, responsible for
comparing two IRs and generating a set of edit operations to transform one
graph into the other. As discussed in the background section, the diff tool
operates based on cost functions for every node and edge operation. To
accurately reflect the nuances of XLS IRs, we implemented custom cost functions
based on a collection of the attributes of nodes and edges that were
precomputed during the parsing stage. These functions are intended to guide the
algorithm toward minimal edit paths while ensuring that the resulting
transformations are feasible and valid within the XLS IR framework, as XLS IR
enforces multiple structural rules that must be respected. The handling of
these constraints is described in more detail in the Patching section.
Table~\ref{tab:cost_attributes} lists the cost attributes used in the diffing
process, grouped into node and edge categories:

\begin{table}[!t]
	\centering
	\caption{Node and Edge Cost Attributes}
	\label{tab:cost_attributes}
	\footnotesize
	\renewcommand{\arraystretch}{1.5}
	\begin{tabular}{| p{0.15\linewidth} | p{0.35\linewidth} | p{0.35\linewidth} |}
		\hline
		\textbf{Type}         & \textbf{Attribute}           & \textbf{Description}                     \\
		\hline
		\multirow{3}{*}{Edge} & \texttt{source\_data\_type}  & Data type of the source node             \\
		                      & \texttt{sink\_data\_type}    & Data type of the sink node               \\
		                      & \texttt{index}               & Position index for non-commutative edges \\
		\hline
		\multirow{5}{*}{Node} & \texttt{op}                  & Operation type of the node               \\
		                      & \texttt{dtype\_str}          & Data type of the node                    \\
		                      & \texttt{operand\_dtype\_str} & Data type of the operand node            \\
		                      & Unique Attribute(s)          & Specific to node type                    \\
		\hline
	\end{tabular}
\end{table}
\begin{itemize}
	\item {\bfseries Edge Cost Attributes:} Although XLS IR does not explicitly
	      represent edges, it models connectivity through node operands and users. When comparing
	      IR edges structurally, data type compatibility between source and sink nodes
	      is crucial. For instance, 4-bit to 8-bit connections cannot be substituted with
	      8-bit to 16-bit ones. The \texttt{source\_data\_type} and \texttt{sink\_data\_type}
	      attributes ensure this compatibility during edge matching.
	      For non-commutative operations, the \texttt{index} attribute preserves operation
	      order since the sink position relative to other inputs must be explicitly tracked
	      for non-commutative operations like concatenation. Only the sink index needs to
	      be captured, while the source index position is not essential for maintaining
	      structural semantics during graph comparison.
	\item {\bfseries Node Cost Attributes:}
	      Nodes also include a standard set of attributes, with additional attributes
	      for specific node types; many nodes, such as literals, have unique
	      attributes, such as a value field, which are added to  their respective
	      unique attributes. These attributes allow the diff tool to effectively
	      capture both common and unique characteristics of nodes, ensuring precise
	      graph comparisons.
\end{itemize}

Table~\ref{tab:cost_values} summarizes the costs applied during the diff
process. The substitution cost function assigns an infinite cost to mismatched
substitutions ensuring that the algorithm only opts for substitutions when cost
attributes align. In other words, when a node or edge does not match its
counterpart, the tool will find it more cost-effective to delete and then
re-insert the node or the edge with the expected attributes, rather than a
forced substitution with infinite cost. Meanwhile, insertions and deletions are
given a uniform cost of 1, providing a consistent baseline for modifications
and resulting in a simplified decision-making landscape.
\begin{table}[!t]
	\centering
	\caption{Cost Assignment for Graph Edit Paths}
	\label{tab:cost_values}
	\footnotesize
	\begin{tabular}{| p{0.15\linewidth} | p{0.50\linewidth} | p{0.15\linewidth} |}
		\hline
		\textbf{Type}         & \textbf{Operation}                  & \textbf{Cost} \\
		\hline
		\multirow{4}{*}{Edge} & Substitution (identical attributes) & 0             \\
		                      & Substitution (different attributes) & $\infty$      \\
		                      & Insertion                           & 1             \\
		                      & Deletion                            & 1             \\
		\hline
		\multirow{4}{*}{Node} & Substitution (identical attributes) & 0             \\
		                      & Substitution (different attributes) & $\infty$      \\
		                      & Insertion                           & 1             \\
		                      & Deletion                            & 1             \\
		\hline
	\end{tabular}
\end{table}
\subsection{Patching}
\begin{figure}
	\centering
	\includegraphics[width=0.7\linewidth]{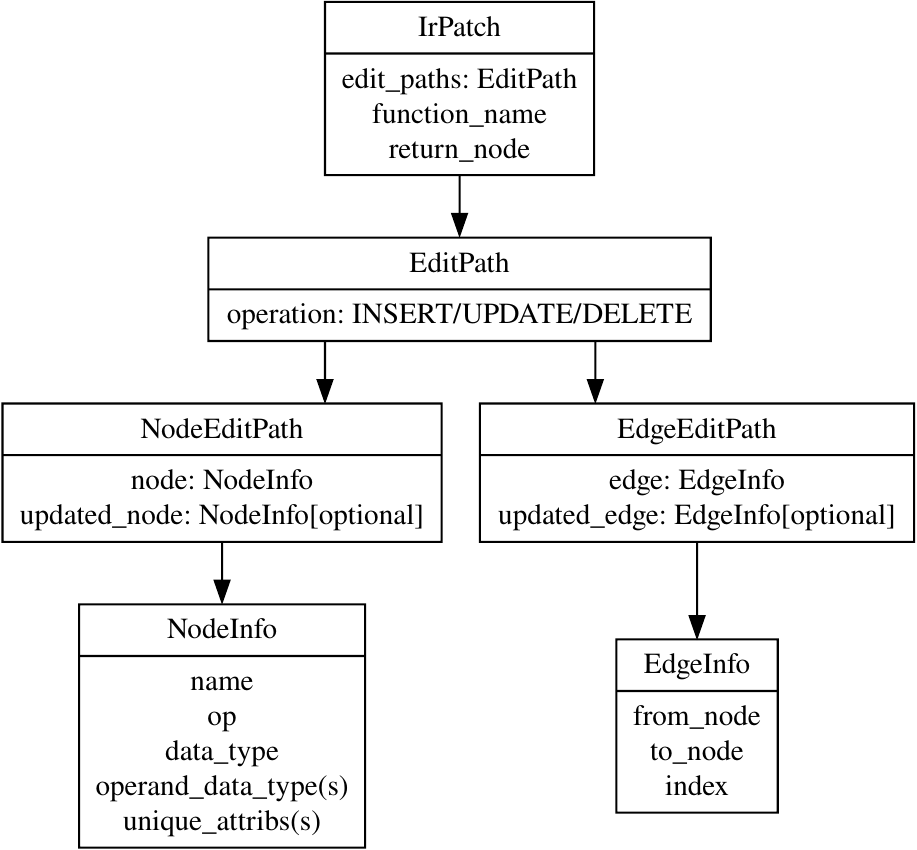}
	\caption{Hierarchical structure of the patch protobuf.}
	\label{fig:patch_structure}

\end{figure}

The patching process enables the systematic transformation of one IR into
another by applying a series of well-defined node and edge insertions,
deletions and substitutions. This process begins by iterating through the
outputs of the IR diff, identifying differences, and capturing them in a
structured format along with the necessary metadata. These differences are then
exported to a protobuf to be consumed by the patch applier later in the flow.
As shown in Figure~\ref{fig:patch_structure}, the protobuf structure is
organized around the IrPatch class, which includes a series of
\texttt{EditPath} operations. Each \texttt{EditPath} operation can pertain to
either a node (\texttt{NodeEditPath}) or an edge (\texttt{EdgeEditPath}) edit
operation. The \texttt{NodeEditPath} and \texttt{EdgeEditPath} further
encapsulate metadata related to nodes (\texttt{NodeInfo}) and edges
(\texttt{EdgeInfo}), respectively.
\subsubsection{Patch Applier}
The patch applier modifies the Intermediate Representation (IR) based on a
sequence of edit instructions, ensuring minimal disruption to the existing IR
structure while maintaining functional correctness. The IR enforces strict
invariants that must be preserved throughout the modification process:
\begin{itemize}
	\item \textbf{Invariant 1:} A node's operands are always populated with valid nodes.
	\item \textbf{Invariant 2:} A node can only be deleted if it has no users.
\end{itemize}
To handle these constraints, the patch applier adopts a staged approach that
prioritizes operations, uses dummy nodes as temporary
stand-ins and leverages dictionaries to track updates, ensuring structural
integrity of the IR during the patching process.

\subsubsection*{Patch Application Order}
The patch application follows a specific order:
\begin{enumerate}
	\item Edge deletions are performed first to disconnect nodes that will later be
	      removed. This ensures safe removal of nodes and allows for edge updates or
	      replacements without conflict. By prioritizing edge deletions, we avoid
	      scenarios where an edge update would subsequently target a deleted edge.
	\item After edge deletions, nodes can be safely removed since they no longer have
	      incoming or outgoing edges, which ensures that the IR remains consistent.
	      During deletion, dummy nodes are used to maintain valid references for
	      dependents, satisfying the first invariant.
	\item Updates and insertions of nodes are prioritized over edge modifications to
	      ensure that all dependencies for edge updates or insertions are already in
	      place. This guarantees that new or updated edges have valid endpoints and
	      dependencies.
	\item Edge modifications are performed last. This guarantees that all nodes are
	      already in their final state, allowing new or updated edges to connect valid
	      endpoints without encountering missing dependencies.
\end{enumerate}

\subsubsection*{Dummy Nodes}
These nodes are used for two primary purposes:
\begin{enumerate}
	\item Before a node is deleted, it is replaced with a dummy node to maintain valid
	      references for its dependents. This ensures that the IR remains structurally
	      sound until the deletion is finalized.
	\item New nodes are initially created with dummy operands that match the
	      required data types. These temporary stand-ins allow the node to exist in a
	      valid state until its final connections are established.
\end{enumerate}
During the patch application, we track which nodes have dummy nodes attached to
them to facilitate proper cleanup. Dummy nodes are not removed or inserted
independently; their lifecycle is tied to the patch operations. When an
insertion operation replaces a dummy node with a real node or when the node
they are attached to is deleted, the dummy nodes are removed automatically.
By the time the patch is fully applied, The IR is expected to contain no dummy
nodes; otherwise, their presence would suggest an error in the process.

\subsubsection*{Replacing Updates with Dictionaries}
Updates are handled without
modifying existing nodes in the IR. When an update requires substituting
a node (e.g., substituting  \texttt{foo} to \texttt{bar}), the patch applier
does not alter original node. Instead, it maintains a dictionary that maps the
new node to the original node. From then on, any reference to \texttt{foo} is
redirected to \texttt{bar}. For changes to the operand index of commutative
operations (where operand order does not affect the operation's result),
the patch applier tracks index changes using a similar  mechanism, which avoids
unnecessary rewiring of nodes and edges. This approach aligns with the primary
goal of the ECO which is minimizing disruption to the existing IR structure while
maximizing reuse.
\subsection{Schedule Constraining:}
The scheduling information in XLS directly determines the RTL structure
produced by the codegen module. Since register names are derived from IR node
names and their cycle assignments, the schedule becomes a crucial bridge
between the IR representation and the final RTL implementation. When applying
ECO, preserving the original schedule becomes essential for maintaining
similarity between the original and revised RTLs. The schedule constraining
phase attempts to maintain the cycle-level scheduling of common nodes in both
versions of the design, while ensuring the resulting schedule remains feasible
under the design's timing constraints. The process begins by examining each
node in the original schedule's cycle map. For each node, we first perform
several qualification checks to determine if the node should be constrained:
\begin{enumerate}
	\item Newly inserted nodes do not have an original placement to preserve, making them
	      exempt.
	\item Nodes that no longer exist in the modified IR cannot be constrained since they
	      have been removed.
	\item Literals are inherently flexible in their timing, and constraints are not
	      applied to them.
\end{enumerate}
For every eligible node, we greedily apply cycle-level constraints, attempting
to place the node in its original pipeline stage. The feasibility of this
constraint is immediately verified by attempting to generate a complete
pipeline schedule. The effectiveness of the constraining process can be
measured by comparing the number of successfully constrained nodes against the
total node count in the design.

\section{Results and Evaluation}
\label{sec:results}

This section presents the evaluation results of the proposed ECO flow applied
to a set of XLS designs, aiming to assess its effectiveness across different
design scales and stages of the ECO flow. Our analysis highlights the
structural reuse achieved through patch generation, the runtime performance of
the graph differencing algorithm, and the effectiveness of schedule
preservation during the ECO flow. All evaluated designs, ECO scripts, and
generated patches can be reproduced from our open-source GitHub
repository~\cite{xls_eco_repo}.
\subsection{Experimental Setup}
All experiments were conducted on a Linux-based system (specifications:
\textit{AMD EPYC 7B13, 64 cores, 128GB RAM, Debian Linux}), and functional
equivalence was verified using Cadence Conformal LEC v22.20. Large designs in
our benchmark set, such as Vector Core and Histogram, caused Python recursion
limit issues when running NetworkX GED. To address this, we manually increased
the recursion threshold. This issue, combined with long runtimes for large
designs, indicate that the current NetworkX implementation is nearing its
scalability boundary and may benefit from future optimizations. As discussed in
the Background section, our approach applies the GED algorithm iteratively. To
maintain practical runtimes, we enforced a maximum time limit of 24 hours per
design.
\subsection{Design Summary and ECO Types}
To comprehensively evaluate the ECO flow across a range of design complexities,
we selected seven XLS designs spanning from small modules (e.g.,
\textit{CRC32}) to large, dataflow-intensive pipelines (e.g., \textit{Vector
	Core}). The node counts range from fewer than 100 to nearly 1000 nodes,
offering a practical upper bound for assessing runtime and memory scalability
within current infrastructure limits, as summarized in
Table~\ref{tab:eco_designs}. Most of these designs are either directly taken
from or adapted from the official examples in the open-source XLS repository.
To fill complexity coverage gaps, two additional designs(\textit{Histogram},
\textit{Vector Core}) were developed specifically for this evaluation. These
custom cases help ensure a more consistent progression across the design
scales.
\begin{table}[!t]
	\centering
	\caption{Experimental ECO Designs Characteristics}
	\label{tab:eco_designs}
	\footnotesize
	\begin{tabular}{| l | c | c | c | c |}
		\hline
		\textbf{Design}    & \textbf{\# Nodes} & \textbf{\# Edges} & \textbf{Depth} & \textbf{Scale} \\
		\hline
		CRC32              & 54                & 74                & 35             & Small          \\
		ZSTD Frame Decoder & 177               & 305               & 28             & Medium         \\
		ApFloat MAC        & 260               & 420               & 41             & Medium         \\
		Simple RISCV       & 418               & 825               & 11             & Medium         \\
		FIR Filter         & 593               & 1037              & 79             & Medium         \\
		Histogram          & 807               & 1460              & 56             & Large          \\
		Vector Core        & 988               & 2022              & 56             & Large          \\
		\hline
	\end{tabular}
\end{table}
The proposed ECO methodology was demonstrated through targeted modifications to
seven distinct XLS designs, encompassing a range of changes from minor
functional tweaks to significant structural rewrites.
Table~\ref{tab:eco_change_schedule} summarizes the ECO changes introduced in
each design and whether they affected the schedule. Notably, in the case of
\textit{ZSTD Frame Decoder}, the modification only altered the schedule by
increasing the number of pipeline stages without changing the DSLX source. As a
result, patch generation for this design is skipped and the flow proceeds directly to schedule constraining.
\begin{table}[!t]
	\centering
	\caption{Summary of Introduced Changes and Their Impact on Scheduling}
	\label{tab:eco_change_schedule}
	\footnotesize
	\begin{tabular}{| m{0.22\linewidth} | m{0.42\linewidth} | m{0.14\linewidth} |}
		\hline
		\textbf{Design}    & \textbf{ECO Change Type}                       & \textbf{Schedule Affected} \\
		\hline
		CRC32              & Polynomial modification                        & No                         \\
		ApFloat MAC        & Increased exponent bit width                   & No                         \\
		Simple RISCV       & Register read optimization                     & No                         \\
		ZSTD Frame Decoder & Increased number of pipeline stages            & Yes                        \\
		FIR Filter         & Refined multiplication by zero handling        & Yes                        \\
		Histogram          & Fixed division by zero in last bin calculation & Yes                        \\
		Vector Core        & Addressed integer overflow                     & Yes                        \\
		\hline
	\end{tabular}
\end{table}
\subsection{Patch Generation and Application}

As outlined in the implementation section, the substitution operations are not
directly applied, instead they are interpreted as preserving structural intent
and therefore considered reused. In our analysis, we classify all non-deleted
nodes and edges, including those involved in substitutions as reused, since
they maintain connectivity and their functional role within the dataflow.
Conversely, changes are defined as the sum of insertions and deletions, which
represent structural modifications. We define the following quantities for each
design. The reuse metric captures preserved structural elements:
\begin{equation}
	\text{Reuse} = (N_{\text{orig}} - N_{\text{del}}) + (E_{\text{orig}} - E_{\text{del}})
	\label{eq:reuse}
\end{equation}
where $N_{\text{orig}}$ and $E_{\text{orig}}$ are the original node and edge counts, and $N_{\text{del}}$ and $E_{\text{del}}$ are the deleted counts.

The change metric quantifies structural modifications:
\begin{equation}
	\text{Change} = N_{\text{add}} + N_{\text{del}} + E_{\text{add}} + E_{\text{del}}
	\label{eq:change}
\end{equation}
where $N_{\text{add}}$ and $E_{\text{add}}$ represent added nodes and edges.

\begin{figure}
	\centering
	\includegraphics[width=\linewidth]{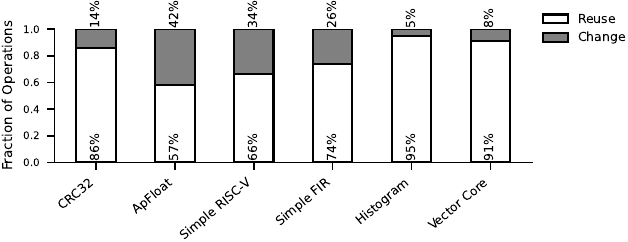}
	\caption{Fraction of reuse (substitutions and unchanged structure) versus structural modification (insertions and deletions) across all evaluated designs. Bars are normalized to total edit operations.}
	\label{fig:reuse_ratios}
\end{figure}

From these, we derive the reuse ratio:
\begin{equation}
	\text{Reuse Ratio} = \frac{\text{Reuse}}{\text{Reuse} + \text{Change}}
	\label{eq:reuse_ratio}
\end{equation}
and the complementary change ratio:
\begin{equation}
	\text{Change Ratio} = 1 - \text{Reuse Ratio}
	\label{eq:change_ratio}
\end{equation}
Using equations (\ref{eq:reuse_ratio}) and (\ref{eq:change_ratio}), Figure~\ref{fig:reuse_ratios} visualizes these ratios across all designs using
a normalized stacked bar chart, while Table~\ref{tab:edit_ops_detailed} lists
the corresponding raw operation counts.
\begin{table}[t]
	\centering
	\caption{Breakdown of node and edge edit operations for each design}
	\label{tab:edit_ops_detailed}
	\begin{tabular}{|l|rrr|rrr|r|}
		\hline
		\multirow{2}{*}{\textbf{Design}}       &
		\multicolumn{3}{c|}{\textbf{Node Ops}} &
		\multicolumn{3}{c|}{\textbf{Edge Ops}} &
		\multirow{2}{*}{\textbf{Cost}}                                                                               \\
		                                       & \textbf{Add} & \textbf{Del} & \textbf{Sub} &
		\textbf{Add}                           & \textbf{Del} & \textbf{Sub} &                                       \\
		\hline
		CRC32                                  & 1            & 1            & 0            & 8   & 8   & 0    & 18  \\
		ApFloat                                & 57           & 56           & 103          & 162 & 162 & 142  & 437 \\
		Simple RISC-V                          & 11           & 127          & 245          & 161 & 391 & 405  & 690 \\
		FIR                                    & 47           & 21           & 420          & 373 & 337 & 571  & 778 \\
		Histogram                              & 3            & 0            & 757          & 96  & 91  & 1334 & 190 \\
		Vector Core                            & 11           & 3            & 845          & 253 & 243 & 1662 & 510 \\
		\hline
	\end{tabular}

\end{table}

The results demonstrate mixed outcomes across different designs. This
variability highlights a critical insight: the effects of small high-level
source modifications on the post-optimized IR structure are difficult to
predict reliably. Table~\ref{tab:edit_ops_detailed} shows the final generated
patches; however, two designs (FIR and Histogram) produced multiple patches
during execution. For FIR, intermediate patches with costs of 784 and 780 were
generated before reaching the final patch cost of 778. Similarly, Histogram
produced intermediate patches with costs of 202 and 196 before achieving the
final cost of 190. The results show that substitution operations outnumber
additions and deletions in most designs, particularly in larger ones. This
confirms that our cost function implementation effectively captures the
structural properties of the XLS IR. It is important to note that our reuse analysis is based on node
count rather than hardware area. Different XLS node types have vastly different
hardware costs; for example, multipliers may represent 80\% of the design area
while constituting only 5\% of the total nodes. Therefore, a low node reuse
ratio does not necessarily indicate poor area efficiency, as the preserved
nodes might include the most area-intensive components.

\begin{figure}
	\centering
	% Panel (a): patch time
	\begin{subfigure}[t]{0.49\textwidth}
		\centering
		\includegraphics[width=\linewidth]{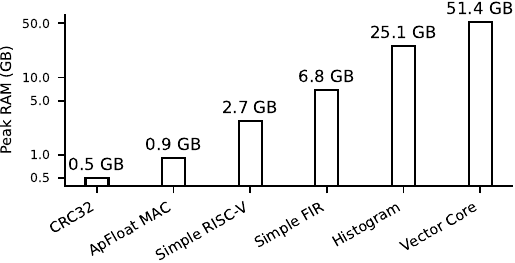}
		\caption{}
		\label{fig:ram_usage}
	\end{subfigure}%
	\hfill
	% Panel (b): RAM usage
	\begin{subfigure}[t]{0.49\textwidth}
		\centering
		\includegraphics[width=\linewidth]{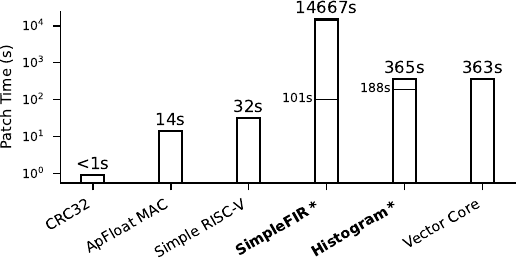}
		\caption{}
		\label{fig:patch_time}
	\end{subfigure}
	\caption{Resource usage during ECO patch generation across designs: (a) Peak RAM consumption (log scale); (b) Total patch generation time (log scale); Designs marked with * yielded more than one patch during the algorithm's execution, with the first patch time shown on their bar.}
	\label{fig:res_usage}

\end{figure}

Figure~\ref{fig:res_usage} summarizes the computational requirements of the
diff process, capturing both peak memory usage and total patch generation time
across designs. As expected, RAM consumption increased with design complexity.
Analysis of peak RAM usage in Figure ~\ref{fig:res_usage}(a) indicates
non-linear memory growth as design complexity increases. While small designs
(\textit{CRC32}, \textit{ApFloat MAC}) required less than 1GB of RAM, larger
ones (\textit{Histogram}, \textit{Vector Core}) exhibited significant
increases, reaching tens of gigabytes. Similar to the memory usage, the patch
generation time, which is shown in Figure~\ref{fig:res_usage}(b), also
demonstrated a non-linear trend, with smaller designs completing in under a
minute, while larger designs (\textit{Histogram}, \textit{Vector Core}) taking
over an hour to generate patches.

\subsection{Schedule Preservation}
\begin{figure}
	\centering
	% Panel (a): patch time
	\begin{subfigure}[t]{0.49\textwidth}
		\centering
		\includegraphics[width=\linewidth]{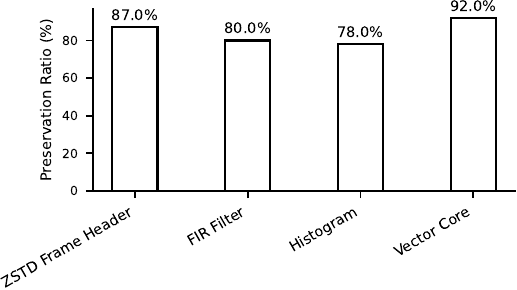}
		\caption{}
		\label{fig:schedule_preservation_ratio}
	\end{subfigure}%
	\hfill
	% Panel (b): RAM usage
	\begin{subfigure}[t]{0.49\textwidth}
		\centering
		\includegraphics[width=\linewidth]{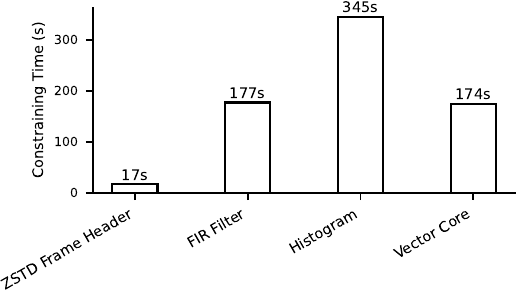}
		\caption{}
		\label{fig:schedule_preservation_time}
	\end{subfigure}
	\caption{Schedule preservation across designs: (a) Percentage of nodes with preserved schedule constraints; (b) Total time taken to resolve schedule
		constraints.}
	\label{fig:schedule_preservation}

\end{figure}

The results presented in Table~\ref{tab:schedule_constraints} and
Figure~\ref{fig:schedule_preservation} highlight the relationship between the
number of structurally infeasible nodes and both the achieved preservation
ratio and the runtime required for constraint resolution. Designs with fewer
infeasible scheduling constraints, such as \textit{ZSTD Frame Decoder}, yielded
high preservation ratios (87\%) and short runtimes (17 seconds). In contrast,
designs exhibiting a higher count of infeasible nodes, such as
\textit{Histogram}, resulted in lower preservation (78\%) and significantly
increased runtimes (345 seconds). The presence of literal-skipped nodes, which
ranged similarly across designs (20–27 nodes), had minimal impact on both
preservation and runtime. The results demonstrate that the proposed scheduling
preservation scheme is effective across a range of design complexities, with
over 75--90\% of original schedule constraints successfully preserved across
all evaluated designs. Given this practical effectiveness, we did not explore
more sophisticated schedule constraining approaches.
\begin{table}[!t]
	\centering
	\caption{Schedule Constraining Summary for Selected Designs}
	\label{tab:schedule_constraints}
	\footnotesize
	\begin{tabular}{| l | c | c | c |}
		\hline
		\multirow{2}{*}{\textbf{Design}} & \multicolumn{3}{c|}{\textbf{Skipped Nodes}}                                                 \\
		\cline{2-4}
		                                 & \textbf{Literal}                            & \textbf{Infeasible} & \textbf{Newly Inserted} \\
		\hline
		ZSTD Frame Decoder               & 25                                          & 1                   & 0                       \\
		FIR Filter                       & 22                                          & 55                  & 0                       \\
		Histogram                        & 27                                          & 93                  & 0                       \\
		Vector Core                      & 13                                          & 46                  & 2                       \\
		\hline
	\end{tabular}
\end{table}

\subsection{Verification}

To evaluate the correctness of the proposed ECO flow, we performed equivalence
checking at both the IR and RTL levels using XLS’s built-in Z3-based
verifier~\cite{xls_ir_equiv} and Cadence Conformal LEC. Among the seven
benchmark designs, five were function-based, and two (\textit{Histogram} and
\textit{ApFloat MAC}) were proc-based. Table~\ref{tab:verification_results}
summarizes the verification outcomes; for function-based designs, IR-level
equivalence was successfully established. Specifically, the equivalence was
verified between the revised IR, generated from the modified design, and the
patched IR, which was derived from the original design by reusing unchanged
components. This confirms that the patching process preserved the intended
functionality. However, since XLS’s Z3-based equivalence checker only supports
function-based designs, IR-level verification could not be applied to
\textit{Histogram} and \textit{ApFloat MAC}, which are proc-based and involve
explicit state and scheduling constructs. To address this, we used Cadence
Conformal LEC for RTL-level verification of the proc-based designs. While
\textit{ApFloat MAC} passed RTL-level equivalence checking, \textit{Histogram}
could not complete verification within a reasonable time due to the design’s
large number of registers and pipeline state elements. In summary, while some
verification limitations remain, especially for designs with complex RTL
states, we validated most benchmarks using the available tools, and the results
confirm the functional correctness of our ECO flow. Future work should explore
more robust verification strategies for handling structural differences in
stateful RTL designs.
\begin{table}[!t]
	\centering
	\caption{Summary of Verification Results}
	\label{tab:verification_results}
	\footnotesize
	\begin{tabular}{| p{0.24\linewidth} | p{0.14\linewidth} | p{0.20\linewidth} | p{0.16\linewidth} |}
		\hline
		\textbf{Design} & \textbf{Type} & \textbf{RTL Equiv.} & \textbf{IR Equiv.} \\
		\hline
		CRC32           & Function      & Passed              & Passed             \\
		ZSTD Decoder    & Function      & Passed              & Passed             \\
		ApFloat MAC     & Proc          & Passed              & N/A                \\
		Simple RISC-V   & Function      & Passed              & Passed             \\
		FIR Filter      & Function      & State mismatch      & Passed             \\
		Histogram       & Proc          & Not completed       & N/A                \\
		Vector Core     & Function      & State mismatch      & Passed             \\
		\hline
	\end{tabular}
\end{table}
\section{Conclusion and future work}
\label{sec:conclusion}

Our evaluation reveals that the proposed approach scales across design
complexities, from small modules to nontrivial dataflow-heavy pipelines, while
maintaining high reuse ratios. The patch applier module, guided by customized
cost functions, enables transformations that respect the structural constraints
of XLS. Applying graph edit sequences in this context requires systematic
updates while preserving functional behavior. Furthermore, the schedule
constraint scheme effectively preserves original timing annotations, helping to
align more internal states in sequential designs and thereby facilitating
equivalence checking in downstream tools and reducing the complexity of later
netlist-level ECOs. Our complete flow, including parser, diff tool, patch
applier, and test designs, is available as an open-source project for the
research community~\cite{xls_eco_repo}.

As for future work, we aim to address several key areas for improvement: (1)
Improving scalability for large designs, the current implementation uses a
single-threaded GED algorithm, which limits performance on complex graphs. The
authors of the core GED algorithm used in our work have also introduced a
multi-threaded variant~\cite{AbuAisheh2018}, which presents a promising
direction for future work. Adapting this parallelized approach to XLS IRs could
substantially enhance runtime performance. Additionally, machine learning-based
approaches could accelerate and scale up the process of graph matching, either
as standalone solutions ~\cite{GEDLIB,GEDGNN,jain2024graph} or as hybrid
approaches that combine classical algorithms with ML
techniques~\cite{Wang_2021_CVPR}. (2) Enhancing verification support, as
mentioned in the verification section, in scenarios where schedules differ,
despite IR-level functional equivalence, conventional sequential equivalence
checking may fail due to mismatched states. In these cases, tools like Synopsys
Formality or hybrid RTL-vs-IR equivalence techniques may offer more robust
validation. (3) Increasing automation, several stages of our flow, such as
feeding diff outputs into the patch applier or identifying schedule divergences
through register comparison, currently require manual effort. Developing
automated pipelines for such tasks would enhance usability and streamline the
ECO process, making it more accessible to users unfamiliar with the underlying
XLS IR structure.

% References
\bibliographystyle{IEEEtran}
\bibliography{ref}

\end{document}